# Calibration of the DHCAL with Muons

**José Repond[1]**

*Argonne National Laboratory*
*Argonne, IL 60439, U.S.A.*
*E-mail:* `repond@hep.anl.gov`

**On behalf of the CALICE collaboration**

**Abstract.** The Digital Hadron Calorimeter (DHCAL) is a large scale prototype of an imaging calorimeter, using Resistive Plate Chambers (RPCs) as active medium. The readout is segmented into $1 \times 1$ cm$^2$ pads, each with a single bit resolution, hence the denomination of digital. The calorimeter counts approximately 500,000 readout channels. The DHCAL construction was completed in spring 2011 and was followed by a series of test beam campaigns in the Fermilab test beam.

In this talk we present the measurement of the performance parameters of the RPCs, i.e. the noise rate, the MIP detection efficiency and the average pad multiplicity. The measurements were performed using trigger-less, as well as triggered data taking in the muon beam of the Fermilab test beam facility. Results are presented on the geometrical alignment, the scan of the response across a single readout pad, and the response over the entire surface of the active layers of the calorimeter. The noise rate measurements identify uncorrelated and correlated noise, as well as contributions from the electronic readout system. When appropriate the measurements are compared to Monte Carlo simulations of the setup.

[1] Speaker

## 1. Introduction

This talk reports on the analysis of muon events collected with the CALICE [1] Digital Hadron Calorimeter (DHCAL) in the Fermilab test beam [2]. Muons produce straight tracks in the DHCAL, which can be used to establish the response of its Resistive Plate Chambers (RPCs), the active media of the calorimeter. The response is characterized by two quantities: the efficiency of detecting Minimum Ionizing Particles (MIPs), ε, and the average pad multiplicity (the number of pads firing for a single MIP traversing the chamber, requiring at least one hit), μ. In addition, the muon tracks provide an excellent tool to align geometrically the individual layers in the DHCAL. The position resolution of the tracks is also sufficient to perform detailed scans of the response across the $1 \times 1$ cm$^2$ readout pads. This talk also reports on measurements of the overall noise or accidental hit rate in the detector.

## 2. Description of the DHCAL

The DHCAL consists of 38 active layers, equipped with RPCs and interleaved with absorber plates. The layers are inserted into the CALICE analog HCAL structure [3], which in turn is placed on a movable stage. Each absorber plate consists of a 17.5 mm thick steel plate and the two covers of the active layers, one being 2 mm thick Copper, the other 2 mm thick Steel. The combined thickness of each absorber layer corresponds to 1.2 radiation lengths $X_0$. A given layer contains three RPCs, each with the dimensions of $32 \times 96$ cm$^2$, and stacked vertiaclly on top of each other. Due to the rims of the RPCs there is a thin inactive region between pairs of chambers in a given layer. The glass plates of an RPC are held apart using four fishing lines with a diameter of 1.15 mm and spaced 5 cm apart.

In addition to the main stack, a Tail Catcher and Muon Tracker (TCMT) structure, located downstream of the DHCAL, is also equipped with RPCs. The TCMT contains 14 active layers. Neglecting the layer covers, the absorber plates for the first 8 layers (last 6 layers) are 19 (102) mm thick.

The RPCs are read out with large Readout boards, consisting of a Pad- and a Front-end board and placed directly on the chambers. Each $1 \times 1$ m$^2$ layer is seen by six readout boards, each with the dimensions of $32 \times 48$ cm$^2$, as seen in Fig. 1. There are 1,536 readout pads, each with the dimensions of $1 \times 1$ cm$^2$, per board. Thus a single layer contains 9,216 readout channels. Each array of $8 \times 8$ readout pads is connected to a front-end ASIC, the DCAL III chip. The latter applies a common threshold (variable within the range of a few fC to about 700 fC) to all 64 channels. The data from the DCAL chip consist of a timestamp (with a resolution of 100 ns) and the pattern of hits above a common threshold. The 24 chips of a board are connected to a data concentrator, located on the outer edge of the Front-end board and seen as the dark green area in Fig.1.

During the course of the January 2011 test beam campaign, the Scintillator layers of the Tail Catcher and Muon Tracker (TCMT), located behind the DHCAL stack, were replaced with layers identical to the ones in the DHCAL. By the end of the running period 14 of the 16 slots in the TCMT had been equipped with RPCs. A photograph of the DHCAL stack before cabling is shown in Fig. 2 (left) and of the TCMT in Fig.2 (right). The DHCAL stack contains 114 individual RPCs and 350,208 readout channels. The TCMT adds an additional 129,024 readout channels for a total of 479,232 (a world record in calorimetry). Additional details on the design, construction, and commissioning of the DHCAL and TCMT can be found in [4].

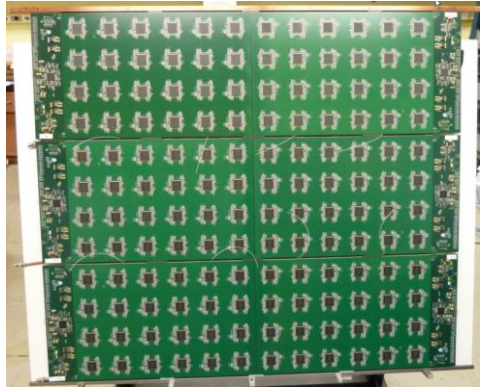

**Figure 1.** Photograph of a DHCAL layer, with the cover removed, showing the top of the six Readout boards.

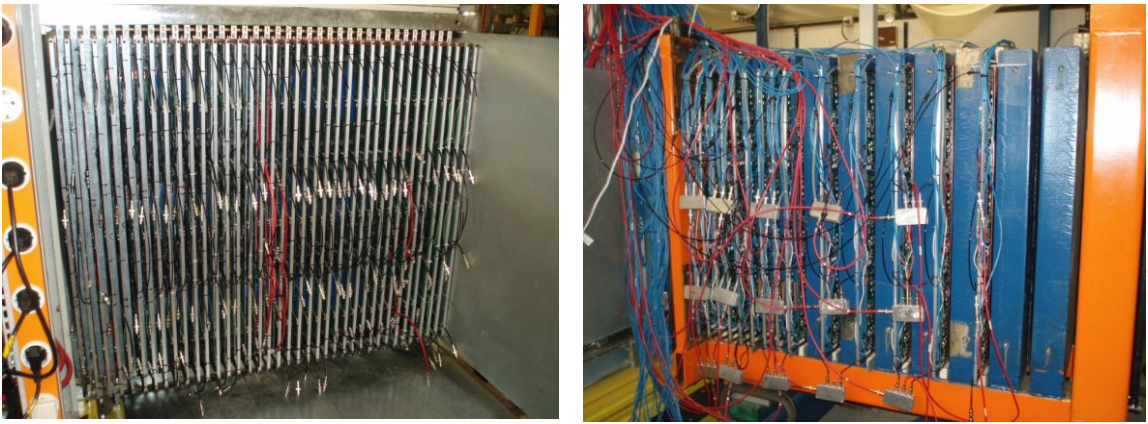

**Figure 2.** Photograph of the DHCAL stack before cabling (left) and of the TCMT (right) equipped with RPCs.

In a digital calorimeter the energy of an incoming particle is reconstructed to first order as being proportional to the number of hits associated with this particle. In the present configuration with $1 \times 1$ cm$^2$ readout pads 1 GeV corresponds to approximately 15 pads [5].

## 3. Trigger and data taking

The stacks were assembled in the Fermilab MTest beam line of the FTBF facility [2]. Broadband muons were generated with the secondary beam tuned to +32 GeV/c and impinging on a 3 m Iron beam blocker placed in the beam line upstream of the experimental area. The data acquisition rates were typically 500 events/spill. The spills lasted approximately 3.5 seconds and were delivered once a minute.

The data acquisition was triggered by two large Scintillator paddles, each with an area of $1 \times 1$ m$^2$, located upstream of the DHCAL stack and behind the TCMT, see Fig.3. The paddle in front was placed about 4.02 ± 0.02 meters upstream of the first absorber plate of the DHCAL. The paddle in the back was implemented into the TCMT structure.

Muon events were collected in several running periods, as shown in Table 1. The numbers of events entered into Table 1 also contain standard CALICE calibration events and a significant fraction of spurious triggers. Therefore, the number of actual muon events is only approximately 65% of the values quoted in the last column. Most of the results presented in this paper are based on the analysis on the data from the October 2010 run.

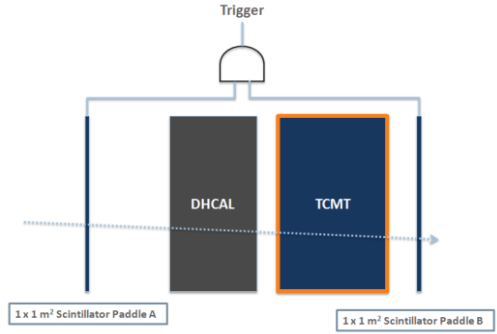

**Figure 3.** Schematic of the set-up for collecting muon events.

**Table 1.** Summary of the various muon runs, indicating the number of active DHCAL and TCMT layers.

| Data taking period | DHCAL layers | TCMT layers | # of events |
|---|---|---|---|
| October 2010 | 38 | 0 | 1,404,700 |
| January 2011 | 38 | 0-13 | 1,583,688 |
| April 2011 | 38 | 14 | 2,483,571 |
| June 2011 | 38 | 14 | 3,286,770 |
| November 2011 | 50 | 0 | 596,674 |
| Total | | | 9,355,403 |

To illustrate the quality of the data taken, Fig. 4 shows two sample events collected with the DHCAL and the 13 layer TCMT. Note the absence of random hits due to noise, which would appear as distinct hits separated from the muon track. Also note that the events in Fig. 4 are shown before application of any noise filtering algorithm.

## 4. Estimation of the noise rate

Noise or accidental hits in the detector will increase the number of hits and thus shift the energy measurement to higher values. It is therefore important to measure and understand the noise rate in great detail. In the DHCAL, the noise rate can be estimated using two separate approaches: i) collecting data in a trigger-less run during a no-beam period. Here all hits together with their time-stamp are collected and recorded and ii) taking a random-trigger run during a no-beam period. Here for a given trigger only hits within a 700 ns time interval are collected and recorded. The following shows results from the first approach. Results from the 2$^{nd}$ approach are in general consistent with the rates presented here, but show some significant differences which can be attributed to some rarely occurring correlated noise hits.

The overall noise rate was observed to be strongly correlated to the temperature of the stack and varied between 0.1 and 1.0 hits/cm$^2$. Table 2 shows the expected contribution to the signal for three typical average noise rates. For a given event trigger hits are recorded in 7 time bins each of 100 ns. However, the subsequent offline analysis only considers hits in the central 2 time bins, thus reducing the effects of possible random noise hits by a factor of 2/7. Note that with a calibration of approximately 15 hits/GeV the noise hits contribute an average of 0.6 to 6.0 MeV (a negligible amount) to the overall event energy.

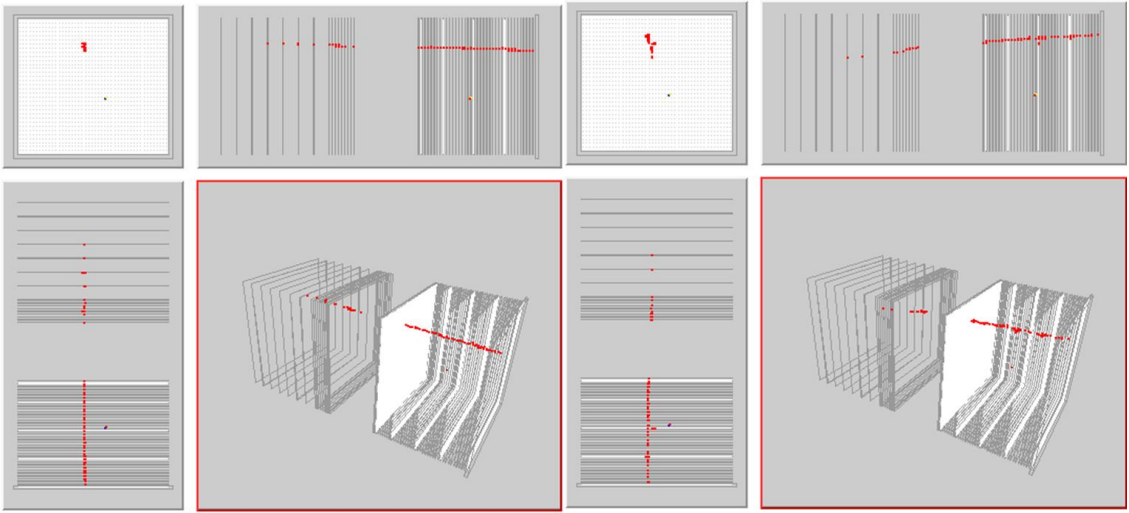

**Figure 4.** Display of two events with single muon tracks traversing the DHCAL and a 13-layer TCMT.

**Table 2.** Expected number of noise hits in the DHCAL + TCMT for three different average noise rates.

| Noise rate [Hz/cm$^2$] | 0.1 | 0.5 | 1.0 |
|---|---|---|---|
| $N_{noise}$/event in DHCAL + TCMT (200 ns) | 0.0094 | 0.047 | 0.094 |
| $N_{noise}$/event in DHCAL + TCMT (700 ns) | 0.033 | 0.165 | 0.33 |

## 5. Reconstruction of muon tracks in the DHCAL

All hits in a given layer are clustered using a closest-neighbor clustering algorithm. The position of a cluster is determined as the unweighted average, $x_{cluster}$ and $y_{cluster}$, of its pads. In order to measure the response in layer i, clusters in all layers (in the following named *tracking layers*), apart from layer i, are used to reconstruct straight tracks. The track reconstruction requires that there be at most one cluster in any tracking layer. Only clusters with at most four hits are accepted for tracking. Finally, for a track to be accepted, it is required to contain tracking clusters in at least 10 layers.

Tracks are reconstructed by performing two independent linear least-square fits in x/z and y/z, respectively. The errors on the cluster positions are assumed to be 1 cm in either lateral coordinate.

A combined, reduced $\chi^2/N_{track}$ of the fits is calculated as

$$\chi^2 / N_{track} = \sum_{j \neq i} \frac{(x^j_{cluster} - x^j_{track})^2}{1} + \sum_{j \neq i} \frac{(y^j_{cluster} - y^j_{track})^2}{1}$$

where the sums run over all tracking clusters, $x^j_{track}/y^j_{track}$ is the inter/extrapolated position of the track in layer j, and $N_{track}$ is the number of clusters used in the fitting. Only tracks with $\chi^2/N_{track}$ smaller than 1.0 are retained for further analysis.

Once a track is reconstructed the track position is inter/extrapolated to layer i, the layer being investigated and not having been used for tracking. Looping over all clusters in layer i, the closest cluster within a radius R = 2.5 cm is considered a match. The number of hits contained in the matching cluster is retained for measuring the pad multiplicity, μ. If no cluster is found within R, the layer is considered inefficient for this track.

With the help of Monte Carlo methods, the resolution of the inter/extrapolated track position was determined to be of the order of 1.3 mm in both x and y.

## 6. Geometrical alignment

The horizontal and vertical alignment of the layers is performed using the residuals, i.e. differences of the inter/extrapolated track and the matching cluster position in both x and y. Figure 5 shows a typical distribution of such residuals. The mean of these residual distributions are used to align the layers. This procedure is seen to work reasonably well and reduces the width of the distribution of the means of the residuals from 1.4 (0.23) mm to 0.57 (0.13) mm in x and y, respectively.

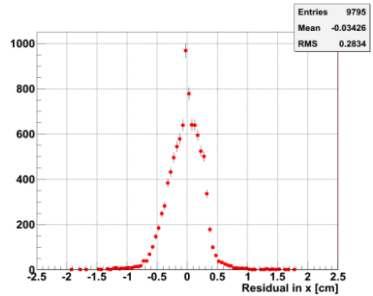

**Figure 5.** Typical distribution of residuals in x.

## 7. Scan across a single pad

Tracks inter/extrapolated into 'clean' regions (see below) are used to scan the response across a $1 \times 1$ cm$^2$ pad. In order to increase the statistics the x/y – positions of all pads are overlaid. Figure 6 shows the average number of hits as function of x, y, and R on the pad for both data and simulation. Here R is defined as

$$R = \sqrt{(x - x_{center})^2 + (y - y_{center})^2}$$

and $x_{center}$ and $y_{center}$ denote the center of the pad. To better illustrate the x(y) dependence, for these plots only tracks in a band of 0.25 < y < 0.75 (0.25 < x < 0.75) have been utilized. The agreement between data and simulation is quite remarkable, given the fact that the present results were obtained by implementing into the simulation the response of RPCs on a first principle basis and not explicitly as function of position on the pad.

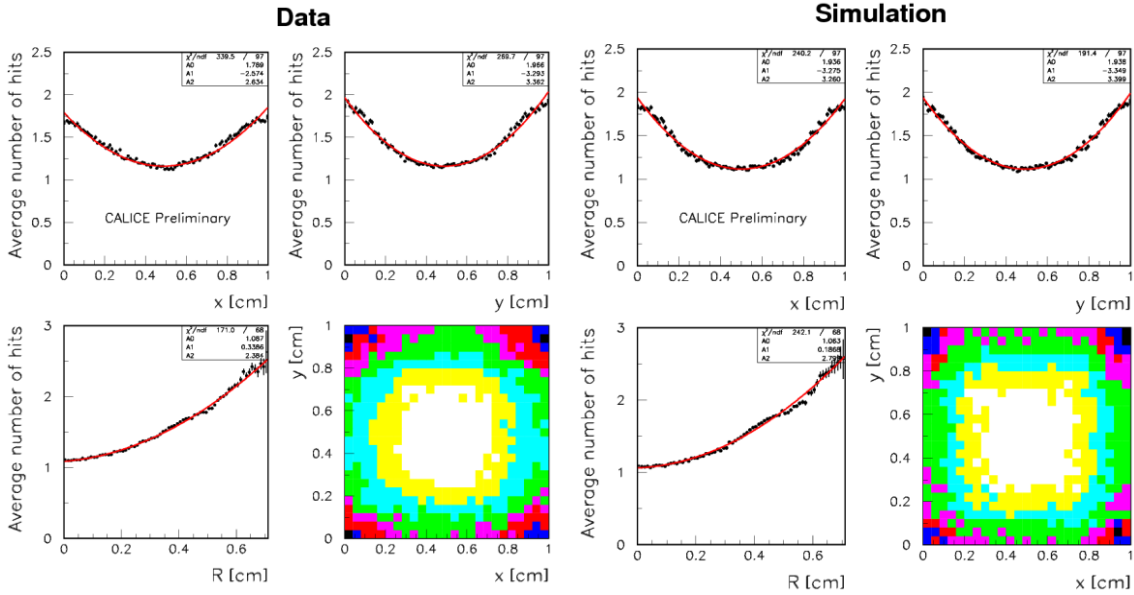

**Figure 6.** Average number of hits (in 'clean' regions of the detector) as function of position on the readout pads: data (left) and simulation (right). The red lines are empirical fits to 2$^{nd}$ second order polynomials.

8. **Response of RPCs**

The response of the RPCs has been measured in 'clean' areas, away from special regions, such as dead ASICs, edges and fishing lines. The list of dead ASICs as a function of time (or run number) was determined using the non-triggered data runs. The data obtained in these 'clean' regions is particularly useful to tune the Monte Carlo simulation of the RPC response.

Figure 7 shows the distribution of the number of pads of all layers combined. The data are displayed as histogram and the simulation as data points. As the momentum distribution of the broadband muon beam is not well known, the momentum of the muons was fixed to 20 GeV/c in the simulation. The results obtained with 10 or 32 GeV/c muons turned out to be indistinguishable. The simulation is seen to adequately reproduce the measured response.

The average response as a function of layer number is shown in Fig. 8. The MIP detection efficiency is seen to be very high, around 94% in the DHCAL. The average pad multiplicity clusters around 1.6, comparable to what was obtained in small scale tests of RPCs at a significantly lower efficiency [6]. The calibration factor $C_i$ for layer i is obtained by multiplying the average efficiency $\varepsilon_i$ and pad multiplicity $\mu_i$ of layer i and dividing by the average values $\varepsilon_0\mu_0$ of the entire stack, $C_i = \varepsilon_i\mu_i/(\varepsilon_0\mu_0)$. For the specific run shown in Fig. 8 the calibration factors are in general above unity, most likely due to elevated temperatures on that day of data taking.

The response of the detector was also measured over the entire surface of a layer. Losses of efficiency are clearly visible at the fishing lines and in the gap between the three RPCs in a given layer. The efficiency, averaged over the entire plane and average over all layers, was measured to be 90.9% with the simulation giving a slightly higher value of 92.1%. The average pad multiplicity was measured to be 1.61, compared to 1.54 in the simulation.

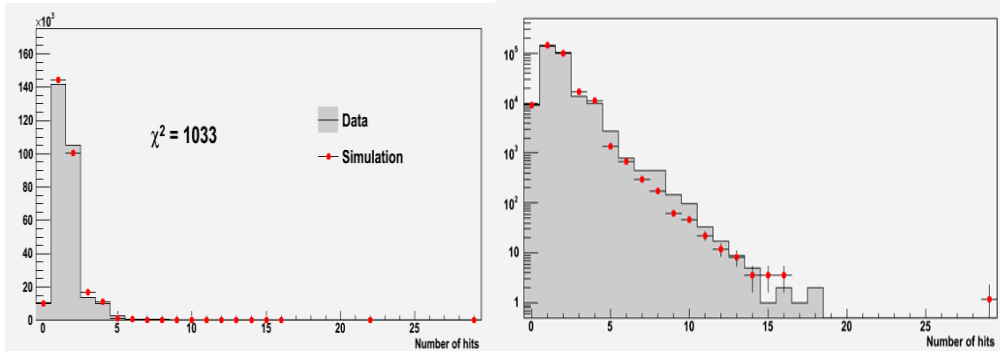

**Figure 7.** Response of RPCs in 'clean' regions, where the histogram (data points) shows the data (simulation): left: linear y-scale, right: logarithmic y-scale.

## 9. Summary

The response of the RPCs in the DHCAL was measured using the Fermilab broadband muon beam. Tracks were reconstructed using hits in all layers, but the one being measured. Based on the residuals, i.e. the difference between a cluster in a given layer and the inter/extrapolated track in that layer the layers were aligned geometrically to 0.57 (0.13) mm in x and y, respectively. Scans across a single pad show a lower pad multiplicities in the center and higher values on the edges, as expected. These effects are well described by the Monte Carlo simulation of the RPC response. The average MIP detection efficiency in 'clean' regions of the RPCs is about 94% and the average pad multiplicity is 1.5 in the DHCAL. After tuning of various parameters in the RPC simulation, these values are well reproduced by the Monte Carlo.

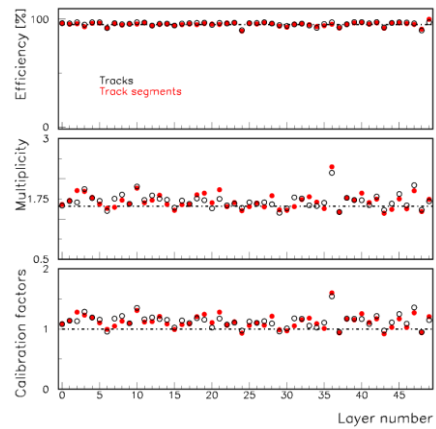

**Figure 8.** Efficiency, average pad multiplicity and calibration factors as function of layer number.